\documentclass[11pt]{article}

\usepackage[preprint]{acl}

\usepackage{times}
\usepackage{latexsym}
\usepackage{amsmath}
\usepackage[T1]{fontenc}
\usepackage{booktabs} 
\usepackage[table]{xcolor}
\usepackage{tabularx} 
\usepackage{multirow}
\usepackage[utf8]{inputenc}

\usepackage{microtype}

\usepackage{inconsolata}

\usepackage{graphicx}
\usepackage{algorithm}
\usepackage{algpseudocode}

\title{Jasper-Token-Compression-600M Technical Report}

\author{Dun Zhang\textsuperscript{\rm 1} \quad Ziyang Zeng\textsuperscript{\rm 2}\thanks{~~Equal contribution.} \\ 
\textbf{Yudong Zhou}\textsuperscript{\rm 1} \quad
\textbf{Shuyang Lu}\textsuperscript{\rm 1} 
\\
\textsuperscript{\rm 1}Prior Shape \\
\textsuperscript{\rm 2}Beijing University of Posts and Telecommunications \\  
\texttt{dunnzhang0@gmail.com, ziyang1060@bupt.edu.cn} \\
\texttt{zhouyudong@priorshape.com, zinkworld@live.cn}
}

\begin{document}
\maketitle
\begin{abstract}
This technical report presents the training methodology and evaluation results of the open-source Jasper-Token-Compression-600M model, released in November 2025. 
Building on previous distillation-based recipes from the English Stella and Jasper models, we successfully extend this approach to a bilingual (English and Chinese) domain, further enhancing model performance through the incorporation of contrastive learning.
A key innovation of our model is the introduction of a one-dimensional convolution-based token compression module. 
We dynamically adjust the compression rate during training, enabling the model to learn more robust and efficient compressed text representations.
By combining knowledge distillation with token compression techniques, we achieve significant improvements in both embedding quality and inference efficiency. 
Our model performs with higher efficiency than a traditional 0.6B model while achieving performance comparable to that of an 8B model.
For more information on the model release, visit: \url{https://huggingface.co/infgrad/Jasper-Token-Compression-600M}.
\end{abstract}

\begin{figure*}[h!]
  \includegraphics[width=\linewidth]{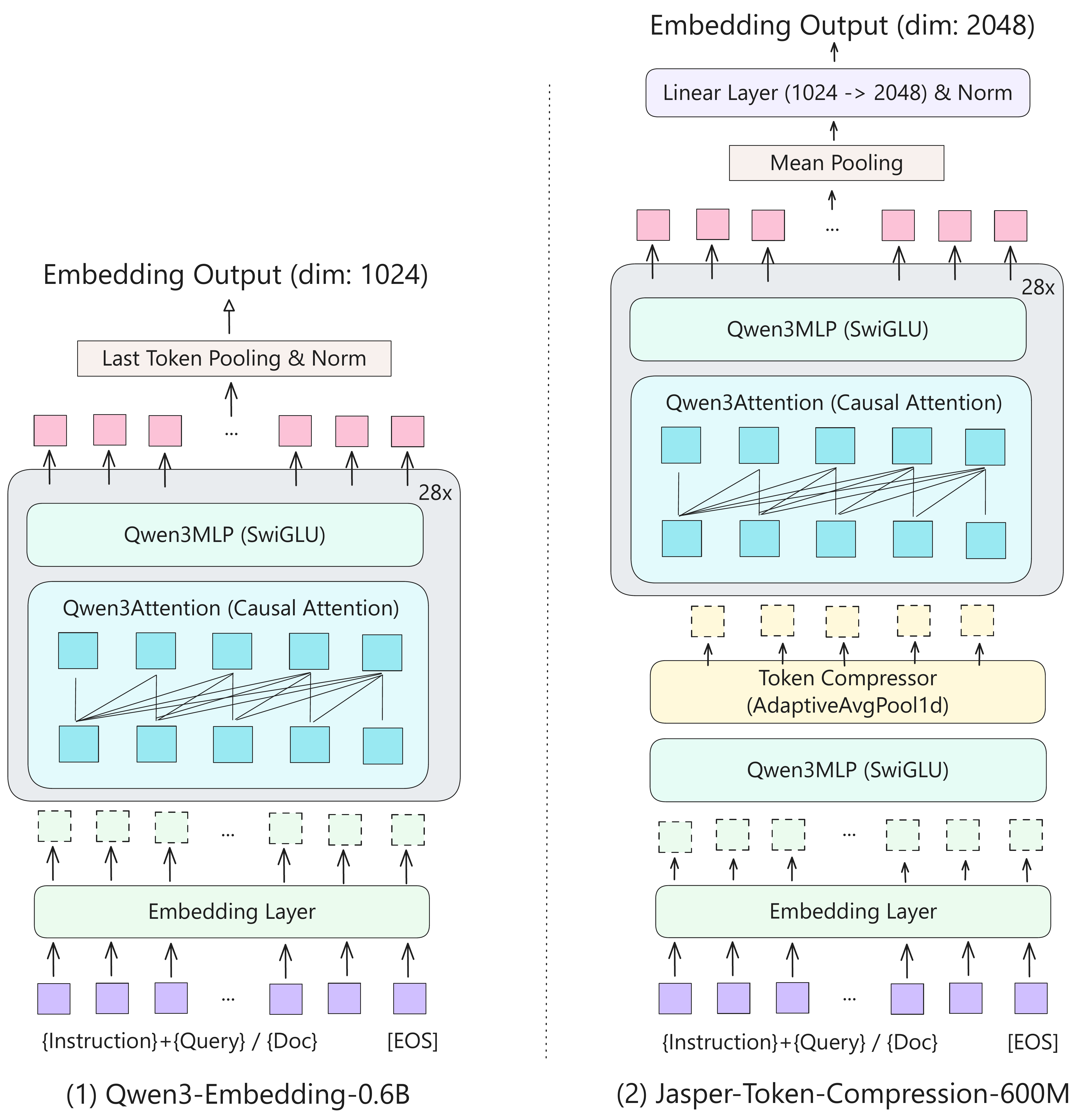}
  \caption{Model architecture of Qwen3-Embedding-0.6B (left) and Jasper-Token-Compression-600M (right).}
  \label{fig:jtc}
\end{figure*}

\section{Introduction}
Text embedding models are key components in natural language processing (NLP) systems.
These models map words, sentences, or documents into a high-dimensional continuous space, where similar texts are represented closer together in vector form, improving the ability to organize, search, and interpret textual data.

However, embedding models that achieve high performance on the MTEB leaderboard\footnote{\url{https://huggingface.co/spaces/mteb/leaderboard}}~\cite{MTEB,enevoldsen2025mmteb} often come with a large number of parameters and high vector dimensions, which result in increased computational and storage demands, making their practical deployment more challenging.
Knowledge distillation is a powerful technique that transfers the capabilities of a large teacher model to a smaller, more efficient student model~\cite{44873}, enabling performance retention while reducing model size. 
This approach has been highly successful in English embedding models, such as Stella and Jasper~\cite{zhang2025jasperstelladistillationsota}.
Despite its success in monolingual settings, the application of distillation in multilingual scenarios remains underexplored. This is an important gap, given the growing need for efficient models that can handle diverse languages.

In this technical report, we present the bilingual Jasper-Token-Compression-600M text embedding model (Figure~\ref{fig:jtc}).
The core of our training approach is knowledge distillation and contrastive learning. Through knowledge distillation on a 12-million-paragraph bilingual text corpus, the 0.6B student model, based on Qwen3-Embedding-0.6B~\cite{zhang2025qwen3embeddingadvancingtext}, learns complementary vector representations from two teacher models: Qwen3-Embedding-8B~\cite{zhang2025qwen3embeddingadvancingtext} and QZhou-Embedding (7B)~\cite{yu2025qzhouembeddingtechnicalreport}. Additionally, contrastive learning further enhances the student model's performance in retrieval tasks.
Inspired by DeepSeek-OCR~\cite{wei2025deepseekocrcontextsopticalcompression}, we introduce a one-dimensional convolution-based token compression module in our embedding model. 
This module applies \texttt{AdaptiveAvgPool1d}\footnote{\url{https://docs.pytorch.org/docs/stable/generated/torch.nn.AdaptiveAvgPool1d.html}} to the input token embeddings, significantly reducing the sequence length fed into the computationally intensive attention module. 
We dynamically adjust the compression rate during training, allowing the model to learn more robust and efficient compressed text representations. 
As a result, our model enables a flexible reduction in input length while preserving strong performance, leading to significant improvements in efficiency.

For model evaluation, we demonstrate that our 0.6B model achieves a Mean(Task) score of 74.75 on the English portion of the MTEB benchmark and a Mean(Task) score of 73.51 on the Chinese portion.
Across multiple compression ratios, Jasper-Token-Compression-600M demonstrates greater efficiency than a traditional 0.6B model, thanks to its token compression module, while delivering performance comparable to that of an 8B model.

\section{Training Methodology}

The training procedure consists of four stages, each focusing on a different capability, as detailed below.

\subsection{Knowledge Distillation}
In the first stage, we apply standard knowledge distillation to help the student model, initialized from Qwen3-Embedding-0.6B, learn the embedding representation capabilities of the teacher models.
We observed that Qwen3-Embedding-8B and QZhou-Embedding (7B) exhibit strong and complementary performance across different tasks on the MTEB leaderboard (e.g., Qwen3-Embedding-8B excels at retrieval tasks (69.44), while QZhou-Embedding performs well on semantic textual similarity tasks (91.65)).
Preliminary experiments show that their embedding representations are indeed complementary, enabling the student model to benefit the most from distillation.\footnote{We also tested combining Qwen3-Embedding-8B with Qwen3-Embedding-4B, but achieved limited improvements.} 
Therefore, we select Qwen3-Embedding-8B and QZhou-Embedding as the two teacher models.

Given that the embedding dimensions of Qwen3-Embedding-8B and QZhou-Embedding are 4096 and 3584, respectively, we apply special processing to facilitate subsequent distillation.
Given that Qwen3-Embedding-8B adopts Matryoshka Representation Learning (MRL)~\cite{10.5555/3600270.3602462} during training, allowing custom output dimensions for the final embedding, we directly take the first 1024 dimensions of its embedding, denoted as $E_{qwen}$.
For QZhou-Embedding which does not support MRL natively, we take the first 3072 dimensions of its 3584-dimensional embedding, split them into three contiguous 1024-dimensional segments, and sum the segments element-wise to obtain a 1024-dimensional vector, denoted as $E_{qzhou}$.
The final 2048-dimensional teacher embedding is then constructed as: $E_t = \text{Norm}( \text{Norm}(E_{\text{qwen}})\, | \,\text{Norm}(E_{\text{qzhou}}))$
, where ``$|$'' denotes concatenation along the feature dimension, and ``Norm'' denotes L2 normalization.
The original Qwen3-Embedding-0.6B produces a 1024-dimensional embedding based on last-token pooling. We modify this to mean pooling and introduce a randomly initialized linear layer to map the 1024-dimensional vector to a 2048-dimensional student embedding. After L2 normalization, the student embedding is denoted as $E_s$.
During this stage, we adopt the classical \emph{cosine loss}, formulated as:
\begin{equation}
 \label{eq:cos_loss}
 \mathcal{L}_{\text{cosine}} = 1 - E_s \cdot E_t .
\end{equation}
The $\mathcal{L}_{\text{cosine}}$  minimizes the angular difference between $E_s$ and $E_t$ in high-dimensional space, thereby aligning their absolute text representations.

\paragraph{Training Setup} We perform distillation training using a 12-million bilingual unsupervised paragraph dataset, with a roughly 1:1 ratio between Chinese and English texts.
During training, the cosine loss is scaled by a factor of 10. 
We set the maximum sequence length to 1,030 tokens and train for 2 epochs using the Adam optimizer with a learning rate of 1e-4, a warmup ratio of 0.005, and cosine learning-rate scheduling. 
Training is conducted on four Nvidia 4090 GPUs with a per-GPU batch size of 4 and 16 gradient accumulation steps, resulting in a global batch size of 256.
We employ FlashAttention-2~\cite{dao2024flashattention} to improve the efficiency of attention computation.

\subsection{Distillation with Token Compression}

In the second stage, we propose a one-dimensional convolution-based token compression module, which is added to our model from the previous stage. 
The text sequence is encoded into token IDs by the tokenizer and input into the model's embedding layer to obtain the embedding vectors. 
At this point, instead of directly passing the embedding vectors into the subsequent attention module, we first introduce a randomly initialized Qwen3MLP (i.e., SwiGLU) Feed-Forward Network to perform feature transformation on the token representations. 
Then, we apply \texttt{AdaptiveAvgPool1d} to perform 1D adaptive average pooling over the token sequence.
\texttt{AdaptiveAvgPool1d} allows us to specify the target sequence length, automatically calculating the kernel size and stride to compress variable-length sequences. 
The algorithm for computing the target sequence length is provided in Algorithm~\ref{alg:target_seq_len}. 
If the target sequence length is set to NULL, no compression is applied. 
By configuring the length threshold and compression ratio, we can flexibly compress long sequences while preserving short ones.
The token compression module shortens the input sequence, thereby reducing the computational burden of the quadratic-complexity attention mechanism and improving overall efficiency.

\paragraph{Training Setup} We perform distillation training using the same data and loss function as in the first stage.
During this stage, all parameters of the student model are trained. It is important to note that the newly introduced token compression module only has the Qwen3MLP as trainable, while the \texttt{AdaptiveAvgPool1d} is training-free.
We train for 2 epochs using the Adam optimizer with a learning rate of 7e-5, with a length threshold of 80 and a compression ratio of 0.33. All other parameters are kept the same as in the first stage.

\begin{algorithm}[t]
\caption{Target Sequence Length Calculation}
\label{alg:target_seq_len}
\begin{algorithmic}[1]
\Require 
    Input length $L_{\mathrm{in}}$, threshold $L_{\mathrm{th}}$, compression ratio $\rho$
\Ensure 
    Target length $L_{\mathrm{tgt}}$

\If{$L_{\mathrm{in}} \leq L_{\mathrm{th}}$}
    \State $L_{\mathrm{tgt}} \gets \text{NULL}$ 
    \Comment{No compression below the threshold}
\Else
    \State $L_{\mathrm{tgt}} \gets 
        L_{\mathrm{th}} + (L_{\mathrm{in}} - L_{\mathrm{th}}) \times \rho$
\EndIf

\end{algorithmic}
\end{algorithm}

\subsection{Distillation with Dynamic Compression}
\label{sec:dc}

In the third stage, we introduce a dynamic compression strategy, where each batch is trained using a randomly sampled compression ratio, instead of the fixed 0.33 ratio used in the second stage. The sampling mechanism is provided in Algorithm~\ref{alg:dynamic_cr}. 
Overall, the model selects the 0.33 compression ratio---consistent with Stage~2---with a probability of 40\%, and explores alternative compression ratios with a probability of 60\%. The goal of this design is to enable the model to maintain stable performance across a wide range of compression intensities, while also allowing flexible compression choices during inference.

In this stage, we augment the original cosine loss $\mathcal{L}_{\text{cosine}}$ with an additional \emph{similarity loss} (defined in Equation~\ref{eq:sim_loss}), which captures semantic alignment between the student and teacher from a pairwise text-similarity perspective.
\begin{equation}
\label{eq:sim_loss}
\mathcal{L}_{\text{similarity}} = \mathrm{MSE}\!\left(BE_{s}BE_{s}^{T},\, BE_{t}BE_{t}^{T}\right)
\end{equation}
Here, $BE_{s}$ and $BE_{t}$ denote the embedding matrices of all texts in the batch produced by the student and teacher models, respectively.
And $\mathrm{MSE}(\cdot)$ denotes the mean squared error, which measures the average squared difference between corresponding entries of two matrices.
The $\mathcal{L}_{\text{similarity}}$ encourages the student model to preserve the teacher model's similarity structure without enforcing exact output matching.
The final loss function in Stage~3 is given by:
\begin{equation}
\label{eq: stage3_loss}
\mathcal{L}_{\text{s3}} = 10 \cdot \mathcal{L}_{\text{cosine}} + 100 \cdot \mathcal{L}_{\text{similarity}}.
\end{equation}

\paragraph{Training Setup}  We train for 800 steps using the Adam optimizer with a learning rate of $7\times10^{-5}$, with a length threshold of 80 and dynamic adjusted compression ratios.
The number of gradient accumulation steps is increased from 16 to 32, resulting in an increase in global batch size from 256 to 512. All other training configurations remain identical to those used in the first stage.

\begin{algorithm}[t]
\caption{Compression Ratio Sampling}
\label{alg:dynamic_cr}
\begin{algorithmic}[1]
\State $r \gets \text{random.random()}$
\If{$r < 0.1$}
    \State $\text{ratio} \gets \text{Uniform}(0.1,\, 0.33)$
\ElsIf{$r < 0.5$}
    \State $\text{ratio} \gets 0.33333$
\ElsIf{$r < 0.8$}
    \State $\text{ratio} \gets \text{Uniform}(0.33,\, 0.66)$
\Else
    \State $\text{ratio} \gets \text{Uniform}(0.66,\, 1.0)$
\EndIf
\end{algorithmic}
\end{algorithm}

\subsection{Contrastive Learning for Retrieval}
\label{sec:cl}
After the three-stage distillation procedure described above, we observed that the student model achieves performance comparable to the teacher models on most MTEB tasks. 
However, its retrieval performance still lag behind those of the teacher models. 

To address this gap, we introduce the fourth contrastive learning-based training stage, specifically designed to enhance retrieval ability.
We adopt an improved contrastive objective built upon the InfoNCE framework~\cite{oord2019representationlearningcontrastivepredictive}. Given a batch of $N$ training instances, the \emph{contrastive loss} is defined as:
\begin{equation}
\label{eq:retrieval}
    \mathcal{L}_{\text{cl}} = - \frac{1}{N} \sum_{i=1}^{N} 
    \log \frac{\exp\left(s(q_i, d_i^+) / \tau\right)}{Z_i},
\end{equation}
where $s(\cdot, \cdot)$ denotes the similarity function (cosine similarity in our setting), $\tau$ is a temperature parameter, and $Z_i$ is the normalization term that aggregates the similarity scores of the positive pair against various negative pairs:
\begin{equation}
\begin{aligned}
Z_i =\;& 
    \exp\!\left(s(q_i, d_i^+)/\tau\right) \\
    &+ \sum_{k=1}^{K} 
        \exp\!\left(s(q_i, d_{i,k}^-)/{\tau}\right) \\
    &+ \sum_{j \neq i} 
        \exp\!\left(s(q_i, d_j)/{\tau}\right).
\end{aligned}
\end{equation}
Here, these terms respectively correspond to similarities with:
(1) the positive document $d_i^+$;
(2) the $K$ hard negative documents $d_{i,k}^-$;
(3) the remaining in-batch documents $d_j$ contrasted with the query $q_i$, yielding $(N - 1)(1 + K)$ easy negatives.

\begin{table*}[ht]
\centering
\begin{tabular}{l c c c c}
\toprule
\textbf{Model} & \textbf{Size} & \textbf{Dimension} & \textbf{Mean (Task)} & \textbf{Mean (Task Type)} \\
\midrule
stella\_en\_400M\_v5 & 435M & 4096 & 69.39 & 64.84 \\
KaLM-embedding-mini-v2.5 & 494M & 896 & 71.29 & 65.31 \\
\rowcolor{black!10}
Qwen3-Embedding-0.6B & 595M & 1024 & 70.70 & 64.88 \\
\rowcolor{red!10}
Jasper-Token-Compression-600M & 600M & 2048 & 74.75 & 68.46 \\
jasper\_en\_vision\_language\_v1 & 1.5B & 8960 & 71.41 & 66.65 \\
F2LLM-1.7B & 1.7B & 2560 & 72.01 & 65.67 \\
Qwen3-Embedding-4B & 4B & 2560 & 74.60 & 68.10 \\
F2LLM-4B & 4B & 2560 & 73.67 & 67.29 \\
\rowcolor{blue!10}
QZhou-Embedding & 7B & 3584 & 75.97 & 69.52 \\
LGAI-Embedding-Preview & 7B & 4096 & 74.12 & 68.40 \\
Linq-Embed-Mistral & 7B & 4096 & 69.80 & 65.29 \\
SFR-Embedding-Mistral & 7B & 4096 & 69.31 & 64.94 \\
NV-Embed-v2 & 7B & 4096 & 69.81 & 65.00 \\
\rowcolor{blue!10}
Qwen3-Embedding-8B & 8B & 4096 & 75.22 & 68.71 \\
Seed1.5-Embedding & - & 2048 & 74.76 & 68.56 \\
Seed1.6-embedding & - & 2048 & 74.07 & 67.98 \\
gemini-embedding-001 & - & 3072 & 73.30 & 67.67 \\
\bottomrule
\end{tabular}
\caption{Results on the English portion of the MTEB benchmark.
Rows highlighted in blue (QZhou-Embedding and Qwen3-Embedding-8B) correspond to the teacher models,
and the row highlighted in black (Qwen3-Embedding-0.6B) corresponds to the initialized student model.
The model results are taken from the official English MTEB leaderboard.
Full results are provided in Appendix Table~\ref{tab:eng_full_results}.}
\label{tab:english_benchmark}
\end{table*}

\begin{table*}[ht]
\centering
\begin{tabular}{l c c c c}
\toprule
\textbf{Model} & \textbf{Size} & \textbf{Dimension} & \textbf{Mean (Task)} & \textbf{Mean (Task Type)} \\
\midrule
Yinka & 326M & 1024 & 70.70 & 71.73 \\
stella-mrl-large-zh-v3.5-1792d & 326M & 1792 & 68.60 & 69.30 \\
ritrieve\_zh\_v1 & 326M & 1792 & 72.71 & 73.85 \\
xiaobu-embedding-v2 & 326M & 768 & 72.36 & 73.48 \\
Conan-embedding-v1 & 326M & 768 & 72.50 & 73.65 \\
zpoint\_large\_embedding\_zh & 326M & 1792 & 71.81 & 72.82 \\
KaLM-embedding-mini-v2.5 & 494M & 896 & 70.89 & 72.43 \\
\rowcolor{black!10}
Qwen3-Embedding-0.6B & 595M & 1024 & 66.33 & 67.45 \\
\rowcolor{red!10}
Jasper-Token-Compression-600M & 600M & 2048 & 73.51 & 75.00 \\
Youtu-Embedding & 2B & 2048 & 77.58 & 78.86 \\
Qwen3-Embedding-4B & 4B & 2560 & 72.27 & 73.51 \\
QZhou-Embedding-Zh & 7B & 1792 & 78.52 & 80.29 \\
\rowcolor{blue!10}
QZhou-Embedding & 7B & 3584 & 76.99 & 78.58 \\
gte-Qwen2-7B-instruct & 7B & 3584 & 71.62 & 72.19 \\
\rowcolor{blue!10}
Qwen3-Embedding-8B & 8B & 4096 & 73.84 & 75.00 \\

Seed1.5-Embedding & - & 2048 & 74.87 & 76.01 \\
Seed1.6-embedding & - & 2048 & 75.63 & 76.68 \\
Conan-embedding-v2 & - & 3584 & 74.24 & 75.99 \\
piccolo-large-zh-v2 & - & 1024 & 70.86 & 71.94 \\
\bottomrule
\end{tabular}
\caption{Results on the Chinese portion of the MTEB benchmark.
Rows highlighted in blue (QZhou-Embedding and Qwen3-Embedding-8B) correspond to the teacher models,
and the row highlighted in black (Qwen3-Embedding-0.6B) corresponds to the initialized student model.
The model results are taken from the official Chinese MTEB leaderboard.
Full results are provided in Appendix Table~\ref{tab:cn_full_results}.}
\label{tab:chinese_benchmark}
\end{table*}

To effectively guide the student model during contrastive learning, we further incorporate a \emph{soft distillation loss}:
\begin{equation}
\label{eq:soft}
    \mathcal{L}_{\mathrm{soft}} = 
    \mathrm{D_{KL}}\!\left(
        \mathrm{sm}\!\left(S^{(s)}/{\alpha}\right)
        \Big\|
        \mathrm{sm}\!\left(S^{(t)}/{\alpha}\right)
    \right),
\end{equation}
where $S^{(s)}$ and $S^{(t)}$ denote the similarity score vectors computed from the student and teacher embeddings (each of length $N(1+K)$), respectively, $\mathrm{sm}(\cdot)$ denotes the softmax function, and $\alpha$ is a temperature scaling factor used to adjust similarity differences.
Additionally, we employ the $\mathcal{L}_{\text{cosine}}$ previously used in the three training stages as a supplementary regularization term.
The final loss function for Stage 4 is therefore:
\begin{equation}
    \mathcal{L}_{\text{s4}} = \mathcal{L}_{\text{cl}} + 16 \cdot \mathcal{L}_{\text{soft}} + 10 \cdot \mathcal{L}_{\text{cosine}}.
\end{equation}

\paragraph{Training Setup}
We adopt the retrieval training datasets used by Qzhou Embedding. 
Training is conducted for 5{,}000 steps using the Adam optimizer with a learning rate of $2\times10^{-5}$, 
a length threshold of 80, and dynamically adjusted compression ratios. 
We set the hyperparameters $N$ and $K$ to 16 and 3, respectively, and set $\tau$ and $\alpha$ to 0.3 and 0.1, respectively.
We apply a gradient accumulation step of 1 and train across four GPUs with gradient checkpointing enabled.

\section{Experiments}
\subsection{Main Results}
\paragraph{English Text Embedding Benchmark}
We evaluate our model on the English portion of the MTEB benchmark, using a length threshold of 80 and a compression ratio of 0.5 during evaluation. As shown in Table~\ref{tab:english_benchmark}, our proposed Jasper-Token-Compression-600M model achieves strong performance despite its relatively small size. Compared to the initialized student model, Qwen3-Embedding-0.6B, it improves the Mean(Task) score from 70.70 to 74.75 and the Mean(TaskType) score from 64.88 to 68.46, yielding gains of +4.05 and +3.58, respectively. This performance is comparable to, and in some cases competitive with, much larger 4B–8B embedding models. These consistent improvements demonstrate the effectiveness of our token-compression distillation strategy in the English domain.

\paragraph{Chinese Text Embedding Benchmark}
We evaluate our model on the Chinese portion of the MTEB benchmark, using a length threshold of 80 and a compression ratio of 0.5 during evaluation. As shown in Table~\ref{tab:chinese_benchmark}, our proposed Jasper-Token-Compression-600M model demonstrates impressive performance despite its relatively small size. When compared with the initialized student model, Qwen3-Embedding-0.6B, it improves the Mean(Task) score from 66.33 to 73.51 and the Mean(TaskType) score from 67.45 to 75.00, resulting in gains of +7.18 and +7.55, respectively. These improvements place Jasper-Token-Compression-600M in close competition with much larger models, such as Qwen3-Embedding-8B.
The consistent performance improvements across both English and Chinese domains highlight the effectiveness of our bilingual distillation strategy.

\subsection{Ablation Study}

\paragraph{The Effectiveness of Contrastive Learning}
As discussed in Section~\ref{sec:cl}, we introduce the fourth training stage based on contrastive learning to specifically enhance the student model’s retrieval capability. As shown in Table~\ref{tab:cl}, the Stage~3 student model already matches the teacher model (Qwen3-Embedding-8B) on tasks such as Classification and Clustering. However, it still lags behind the teacher by nearly 4 points on asymmetric Retrieval tasks (65.53 vs. 69.44).
After applying the retrieval-oriented contrastive learning in Stage~4, we observe a modest improvement on the Retrieval task (from 65.33 to 66.19), while performance on other tasks fluctuates only slightly. 
This result indicates that targeted contrastive learning is indeed effective for improving retrieval capability.

\begin{table}[ht!]
% \centering
\begin{tabular}{lccc}
\toprule
\multicolumn{1}{l}{\textbf{Task Type}} & \multicolumn{1}{c}{\textbf{Stage~3}} & \multicolumn{1}{c}{\textbf{Stage~4}} & \multicolumn{1}{c}{\textbf{QE8}} \\
\midrule
Classification & 90.49 & 90.35 & 90.43 \\
Clustering & 59.71 & 59.44 & 58.57 \\
PairClassification & 90.08 & 90.15 & 87.52 \\
Reranking & 50.84 & 50.60 & 51.56 \\
\rowcolor{black!10}
Retrieval & 65.53 & 66.19 & 69.44 \\
STS & 88.73 & 88.79 & 88.58 \\
Summarization & 33.28 & 33.66 & 34.83 \\
\midrule
Mean(Task Type) & 68.38 & 68.46 & 68.71 \\
Mean(Task) & 74.65 & 74.75 & 75.22 \\
\bottomrule
\end{tabular}
\caption{Results of Stage~3 (distillation) and Stage~4 (contrastive learning) on the English portion of the MTEB benchmark.``QE8'' refers to Qwen3-Embedding-8B.}
\label{tab:cl}
\end{table}

\begin{table*}[!ht]
\centering
\begin{tabular}{lcccccc}
\toprule
\multirow{2}{*}{\textbf{Model}} & \multirow{2}{*}{\textbf{Mean (Task)}} & \multicolumn{5}{c}{\textbf{Input Length}} \\
\cmidrule(lr){3-7}
 &  & 128 & 256& 512& 1024 & 2048 \\
 \midrule
Qwen3-Embedding-0.6B & 70.70 & 3.22 & 6.47 & 12.20 & 24.24 & 49.99 \\
Jasper-Token-Compression-600M-0.50× & 74.75 & 2.62 & 3.96 & 7.39 & 13.11 & 25.07 \\
Jasper-Token-Compression-600M-0.33× & 74.70 & 2.52 & 3.52 & 5.41 & 9.38 & 17.52 \\
Jasper-Token-Compression-600M-0.20× & 74.58 & 2.38 & 2.91 & 4.00 & 6.56 & 11.48 \\
Jasper-Token-Compression-600M-0.10× & 74.21 & 2.09 & 2.56 & 3.18 & 4.48 & 6.95 \\
\bottomrule
\end{tabular}
\caption{Comparison of MTEB Mean(Task) performance and encoding latency under different token compression ratios. ``Jasper-Token-Compression-600M-0.50×'' denotes a 0.5 compression ratio, and similarly for the other variants.}
\label{tab:ratio}
\end{table*}

\paragraph{The Influence of Token Compression}
As discussed in Section~\ref{sec:dc}, we introduce a compression-ratio sampling mechanism during training to expose the model to a wide range of compression scenarios. This enables flexible compression-ratio selection during inference.
To analyze the impact of different compression ratios on model performance, we experiment with four settings: 0.5 (default), 0.33, 0.2, and 0.1.
Table~\ref{tab:ratio} reports the Mean(Task) scores on the MTEB benchmark under these compression ratios.
To further evaluate how compression improves model efficiency, we measure encoding latency on 1,600 fixed-length texts using a batch size of 32. 
The input length varies from 128 to 2028 tokens. 
% We use the \texttt{SentenceTransformers.encode} function\footnote{\url{https://sbert.net/}} implemented with FlashAttention-2, and compute the average per-sample encoding time (in milliseconds).
We compute the average per-sample encoding time (in milliseconds).
Table~\ref{tab:ratio} shows how encoding time scales with input length across different compression ratios, illustrating that lower compression ratios consistently lead to notable latency reductions while maintaining competitive performance.

\section{Conclusion}
In this technical report, we present Jasper-Token-Compression-600M, a bilingual text-embedding model trained through a multi-stage pipeline. By publicly releasing the model weights, we aim to empower practitioners to use this efficient and effective model for information retrieval, semantic similarity, and clustering tasks across both English and Chinese domains.

\section*{Limitations}
\paragraph{Retrieval Performance}
Despite our efforts to incorporate contrastive learning to enhance the student model’s retrieval capability, a noticeable performance gap remains between the student and teacher models (66.19 vs. 69.44). This suggests that our current contrastive learning setup is still insufficient. Further improving retrieval performance in distilled embedding models remains a promising direction for future work.

\paragraph{Token Compression}
Our current approach uses a simple, training-free one-dimensional convolution–based compression mechanism. While efficient, this design may limit the model’s full potential in text compression. Future work should explore adaptive and trainable compression strategies that adjust to factors such as input length and batch size, enabling a deeper investigation into the role of token compression in embedding models.

\paragraph{Supported Text Length}
The model was distilled using training samples with a maximum length of 1,030 tokens. As a result, performance may degrade when handling longer texts, limiting the model’s capability for long-text representation. Extending distillation to substantially longer sequences is an important direction for future improvement.

\bibliography{custom}

\appendix

\section{Appendix}
\label{sec:appendix}

\begin{table*}[ht]
\centering
\begin{tabular}{llc}
\toprule
\multicolumn{1}{l}{\textbf{Task Type}} & \multicolumn{1}{l}{\textbf{Testset}} & \multicolumn{1}{c}{\textbf{Jasper-Token-Compression-600M}} \\
\midrule
Classification & AmazonCounterfactualClassification & 93.52 \\
Classification & Banking77Classification & 87.46 \\
Classification & ToxicConversationsClassification & 91.06 \\
Classification & MassiveIntentClassification & 85.29 \\
Classification & MassiveScenarioClassification & 90.85 \\
Classification & TweetSentimentExtractionClassification & 78.57 \\
Classification & MTOPDomainClassification & 98.97 \\
Classification & ImdbClassification & 97.11 \\
Clustering & StackExchangeClusteringP2P.v2 & 53.54 \\
Clustering & MedrxivClusteringS2S.v2 & 46.23 \\
Clustering & ArXivHierarchicalClusteringP2P & 66.10 \\
Clustering & TwentyNewsgroupsClustering.v2 & 67.61 \\
Clustering & BiorxivClusteringP2P.v2 & 51.96 \\
Clustering & ArXivHierarchicalClusteringS2S & 63.87 \\
Clustering & MedrxivClusteringP2P.v2 & 49.01 \\
Clustering & StackExchangeClustering.v2 & 77.23 \\
PairClassification & TwitterURLCorpus & 88.91 \\
PairClassification & SprintDuplicateQuestions & 98.34 \\
PairClassification & TwitterSemEval2015 & 83.20 \\
Reranking & MindSmallReranking & 32.77 \\
Reranking & AskUbuntuDupQuestions & 68.44 \\
Retrieval & TRECCOVID & 89.92 \\
Retrieval & Touche2020Retrieval.v3 & 67.94 \\
Retrieval & SCIDOCS & 25.51 \\
Retrieval & HotpotQAHardNegatives & 76.67 \\
Retrieval & CQADupstackUnixRetrieval & 59.13 \\
Retrieval & CQADupstackGamingRetrieval & 70.85 \\
Retrieval & FiQA2018 & 53.89 \\
Retrieval & FEVERHardNegatives & 93.66 \\
Retrieval & ClimateFEVERHardNegatives & 46.03 \\
Retrieval & ArguAna & 78.28 \\
STS & STS13 & 93.35 \\
STS & STS14 & 89.63 \\
STS & STS12 & 86.07 \\
STS & STS17 & 93.42 \\
STS & STSBenchmark & 92.89 \\
STS & STS22.v2 & 73.30 \\
STS & SICK-R & 85.30 \\
STS & BIOSSES & 91.68 \\
STS & STS15 & 93.47 \\
Summarization & SummEvalSummarization.v2 & 33.66 \\
\bottomrule
\end{tabular}
\caption{Results for each dataset in the English portion of the MTEB benchmark.}
\label{tab:eng_full_results}
\end{table*}

\begin{table*}[ht]
\centering
\begin{tabular}{llc}
\toprule
\multicolumn{1}{l}{\textbf{Task Type}} & \multicolumn{1}{l}{\textbf{Testset}} & \multicolumn{1}{c}{\textbf{Jasper-Token-Compression-600M}} \\
\midrule
Classification & JDReview & 88.39 \\
Classification & OnlineShopping & 94.56 \\
Classification & TNews & 57.04 \\
Classification & MultilingualSentiment & 81.23 \\
Classification & Waimai & 90.05 \\
Classification & IFlyTek & 55.03 \\
Clustering & CLSClusteringP2P & 69.94 \\
Clustering & CLSClusteringS2S & 67.98 \\
Clustering & ThuNewsClusteringS2S & 84.07 \\
Clustering & ThuNewsClusteringP2P & 87.82 \\
PairClassification & Ocnli & 83.00 \\
PairClassification & Cmnli & 87.76 \\
Reranking & CMedQAv2-reranking & 88.59 \\
Reranking & T2Reranking & 67.54 \\
Reranking & CMedQAv1-reranking & 88.44 \\
Reranking & MMarcoReranking & 35.24 \\
Retrieval & VideoRetrieval & 76.10 \\
Retrieval & MMarcoRetrieval & 83.93 \\
Retrieval & EcomRetrieval & 68.93 \\
Retrieval & CmedqaRetrieval & 47.54 \\
Retrieval & T2Retrieval & 86.14 \\
Retrieval & CovidRetrieval & 87.52 \\
Retrieval & MedicalRetrieval & 65.37 \\
Retrieval & DuRetrieval & 89.54 \\
STS & QBQTC & 47.04 \\
STS & AFQMC & 56.07 \\
STS & ATEC & 53.87 \\
STS & LCQMC & 80.68 \\
STS & STSB & 88.66 \\
STS & BQ & 72.64 \\
STS & PAWSX & 48.17 \\
\bottomrule
\end{tabular}
\caption{Results for each dataset in the Chinese portion of the MTEB benchmark.}
\label{tab:cn_full_results}
\end{table*}

\end{document}